\documentclass[fleqn,usenatbib]{mnras}

\usepackage{newtxtext,newtxmath}

\usepackage[T1]{fontenc}

\DeclareRobustCommand{\VAN}[3]{#2}
\let\VANthebibliography\thebibliography
\def\thebibliography{\DeclareRobustCommand{\VAN}[3]{##3}\VANthebibliography}

\usepackage{graphicx}	
\usepackage{amsmath}	

\title[Search for planets around Hyades white dwarfs]{Search for giant planets around seven white dwarfs in the Hyades cluster with the Hubble Space Telescope}

\author[W. Brandner, H. Zinnecker and T. Kopytova]{
Wolfgang Brandner,$^{1}$\thanks{E-mail: brandner@mpia.de}
Hans Zinnecker$^{2,3}$
and Taisiya Kopytova$^{4,5,1}$
\\
$^{1}$Max-Planck-Institut f\"ur Astronomie, K\"onigstuhl 17, 69117 Heidelberg, Germany\\
$^{2}$Deutsches SOFIA Institut, Unversit\"at Stuttgart, Germany\\
$^{3}$Universidad Autonoma de Chile, Avda Pedro de Valdivia 425, Santiago de Chile\\
$^{4}$Division of Medical Image Computing, German Cancer Research Center (DKFZ), 69120 Heidelberg, Germany\\
$^{5}$Ural Federal University, Yekaterinburg, 620002, Russia
}

\date{Accepted XXX. Received YYY; in original form ZZZ}

\pubyear{2020}

\begin{document}
\label{firstpage}
\pagerange{\pageref{firstpage}--\pageref{lastpage}}
\maketitle

\begin{abstract}
Only a small number of exoplanets has been identified in stellar cluster environments. We initiated a high angular resolution direct imaging search using the Hubble Space Telescope (HST) and its NICMOS instrument for self-luminous giant planets in orbit around seven white dwarfs in the 625\,Myr old nearby ($\approx$45\,pc) Hyades cluster. The observations were obtained with NIC1 in the F110W and F160W filters, and encompass two HST roll angles to facilitate angular differential imaging. The difference images were searched for companion candidates, and radially averaged contrast curves were computed. Though we achieve the lowest mass detection limits yet for angular separations $\ge$0.5$''$, no planetary mass companion to any of the seven white dwarfs, whose initial main sequence masses were $>$2.8\,M$_\odot$, was found. Comparison with evolutionary models yields detection limits of $\approx$5 to 7 Jupiter masses according to one model, and between 9 and $\approx$12\,M$_{\rm Jup}$ according to another model, at physical separations corresponding to initial semimajor axis of $\ge$5 to 8 A.U. (i.e.,\ before the mass loss events associated with the red and asymptotic giant branch phase of the host star).
The study provides further evidence that initially dense cluster environments, which included O- and B-type stars,  might not be highly conducive to the formation of massive circumstellar disks, and their transformation into giant planets (with m$\ge$6\,M$_{\rm Jupiter}$ and a$\ge$6 A.U.). This is in agreement with radial velocity surveys for exoplanets around G- and K-type giants, which did not find any planets around stars more massive than $\approx$3\,M$_\odot$.
\end{abstract}

\begin{keywords}
Planets and satellites: gaseous planets -- Planets and satellites: formation -- Planets and satellites: detection -- Planets and satellites: dynamical evolution and stability -- white dwarfs -- open clusters and associations: individual: Hyades
\end{keywords}




\section{Introduction}

While several 1000 exoplanets and exoplanet candidates have been identified by now \citep{Perryman2018,Shabram2020}, less than 1\% of these reside in stellar clusters \citep{Kovacs14}. Both radial velocity (RV) surveys (e.g., \cite{Paulson04,Guenther05}) and surveys for transiting planets \citep{Gilliland00,vanSaders11,Mann16} in populous open and globular cluster confirm the low frequency of close-in planets around cluster members. The RV discovery of a giant exoplanet with $\approx$7.5\,M$_{\rm Jup}$ in a $\approx$2\,A.U.\ orbit around the 2.7\,M$_\odot$ K0 giant $\epsilon$\,Tau, which is a kinematic member of the Hyades open cluster, led to the suggestion that giant planets around intermediate mass stars do exist \citep{Sato07}.

Direct imaging studies for exoplanets need to overcome both the brightness contrast between a star and its exoplanet, and achieve fine angular resolution in order to separate the exoplanet signal from its host star. Similar to brown dwarfs, planets cool with age and are therefore most easily detectable at young ages while they are still self-luminous in the infrared. Indeed direct imaging searches for exoplanets have been most successful around young stars \citep{Chauvin05,Marois08,Lagrange09,Macintosh15,Chauvin17,Keppler18}. 

With the aim to search for giant planets in an open cluster environment, we selected white dwarfs in the Hyades. 
The Hyades, at an age of $625\pm 50$\,Myr and at an average distance of 45\,pc (e.g., \cite{Perryman98,Kopytova16}), constitute the most nearby open cluster. In addition to 724 stellar systems classified by their proper motion as kinematic member candidates \citep{Roeser11}, seven single white dwarfs (see Table \ref{TargetSample}) and three white dwarfs, which are companions to stars, have been established as Hyades cluster members \citep{Zuckerman87,vonHippel98}.  Prior to GAIA DR2 at least six additional white dwarfs were considered likely members of the Hyades cluster \citep{Schilbach12,Tremblay12,Zuckerman13}. \cite{Salaris2018} confirm two of these (WD0348+339 and WD0400+148) as high probability members.
Born as Herbig Ae stars with initial masses in the range 2.8 to 3.6\,M$_\odot$ \citep{Kalirai14}, the circumstellar disks of the white dwarf progenitors could have been the birthplaces of giant planets. 

White dwarfs offer two advantages for direct imaging surveys for exoplanets \citep{Zinnecker01,Burleigh02,Gould08}. Firstly, planets on circular or moderately eccentric orbits with semimajor axis of several A.U.\ would survive the post-Main Sequence mass loss of the parent star, and would migrate outward adiabatically by a factor equal to the ratio of initial to final stellar mass due to conservation of orbital angular momentum (e.g., \cite{Villaver09,Nordhaus13}). Secondly, because of their small surface area, white dwarfs are considerably less luminous than their early A- or late B-type main sequence progenitors, thus alleviating the contrast requirements. 
The reduced contrast requirements facilitated the first direct imaging detection of a brown dwarf as a companion to the white dwarf GD\,165 \citep{Becklin88}. 

An additional motivation are semianalytic circumstellar disk models by \cite{Kennedy2008}, which predict a linear increase in the occurrence rate of giant planets with stellar mass in the range 0.4 to 3\,M$_\odot$. Direct imaging detections of giant exoplanets orbiting the mid A-type stars HR\,8799 \citep{Marois08,Marois2010} and $\beta$\,Pic \citep{Lagrange2010} might be supportive for these models.
With the exception of WD 0806-661 B \citep{Luhman2011}, direct detection spectroscopic and imaging searches for planetary mass companions to white dwarfs were unsuccessful thus far (e.g.\ \cite{Chu2001,Hogan2011,Xu2015}).
Their proximity with distances between 35 and 50\,pc \citep{BailerJones18}, relatively young age, and main sequence progenitor masses of $\approx$3\,M$_\odot$
make the Hyades cluster white dwarfs promising targets to search for substellar companions at orbital separations of several tens of A.U.


\section{Observations and data reduction}

\begin{figure*}
    \begin{center}
        \includegraphics[width=1.0\textwidth]{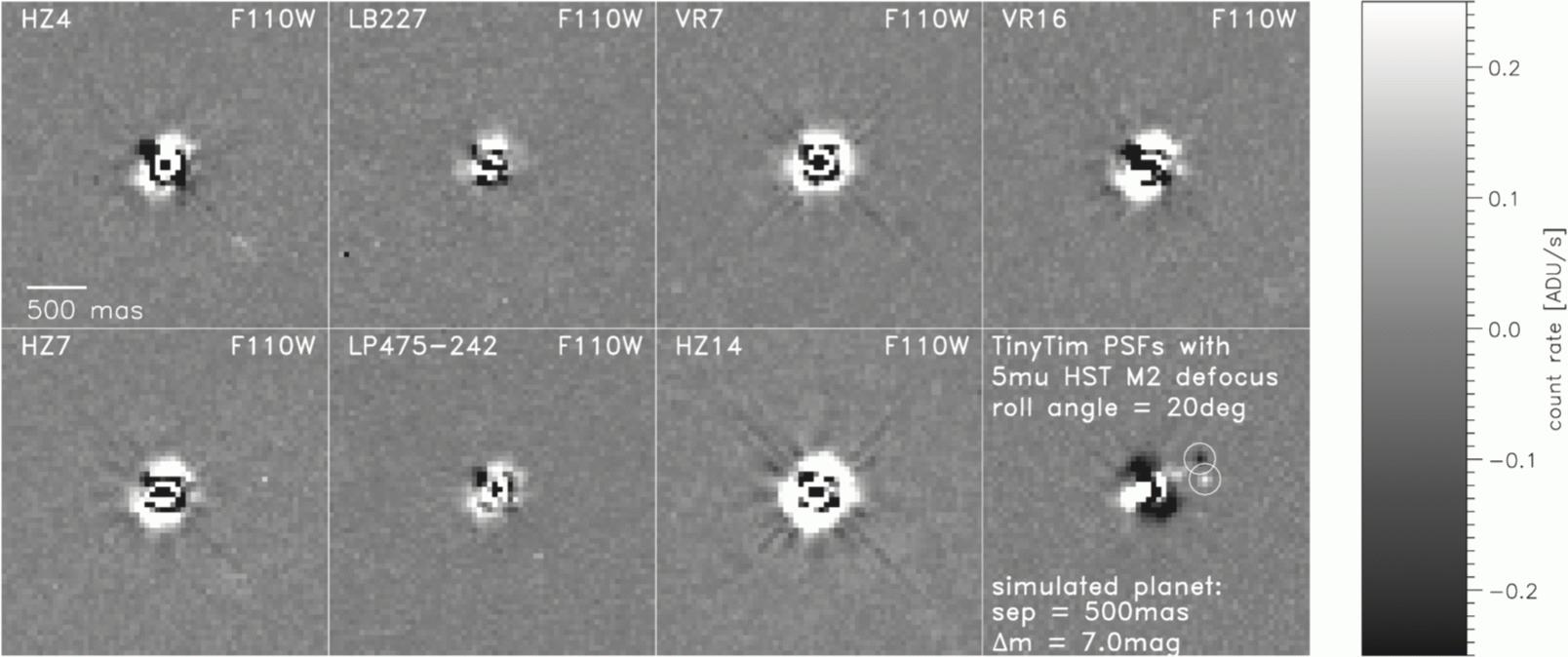}
    \end{center}
    \caption[]{Roll subtracted images of the HST NIC1 observations in F110W of the seven white dwarfs (left to right, top to bottom) and simulated observations based on Tiny Tim PSF simulations \citep{Krist2011} with an exoplanet at a separation of 500 mas, and 7.0 mag fainter than the white dwarf (lower right).}
    \label{NIC1mosaic}
\end{figure*}

\begin{table*}
\caption{Basic astrophysical parameters of the Hyades single white dwarfs and date of HST/NICMOS observations}             
\label{TargetSample}      
\centering                          
\begin{tabular}{l l c c c c c c}        
name & alt.\ name&      J &     H & distance &M$_{\rm init}$  &M$_{\rm final}$ &obs. date \\
     & &        [mag]   &       [mag]   &[pc]&  [M$_\odot$] & [M$_\odot$] & \\ \hline
WD0352+096 &HZ 4 & 14.83$\pm$0.04      &  14.87$\pm$0.06&35.0  &3.59 &0.80  & 2003-11-04\\
WD0406+169 &LB 227     &15.70$\pm$0.07&        15.47$\pm$0.12&50.2     &3.49&0.85 & 2004-02-07\\
WD0421+162& VR 7, LP 415-46    &14.75$\pm$0.04&        14.82$\pm$0.06&45.0     &2.90&0.70 & 2004-02-15\\
WD0425+168& VR 16, LP 415-415& 14.63$\pm$0.03  &       14.65$\pm$0.05&47.9     &2.79&0.71 & 2003-11-05\\
WD0431+126& HZ 7       &14.77$\pm$0.03&                14.80$\pm$0.06&47.3     &2.84&0.69 & 2003-11-06\\
WD0437+138&LP 475-242  &15.32$\pm$0.05 &       15.51$\pm$0.12&46.0     &3.41&0.74 & 2003-11-07\\
WD0438+108& HZ 14      & 14.50$\pm$0.04&       14.62$\pm$0.05&49.4     &2.78&0.73 & 2003-11-09\\
\hline                                   
\end{tabular}
     \begin{quote}
        Notes: apparent magnitudes are from 2MASS; distances are based on GAIA DR2 parallaxes (\cite{GAIA2016,Brown18,BailerJones18}), which within the uncertainties are in very good agreement with the distances previously reported by \cite{Schilbach12}; initial and final mass estimates are from \cite{Kalirai14}
      \end{quote}
\end{table*}

The data were obtained with HST/NICMOS (GO 9737, PI H.\ Zinnecker).  Table \ref{TargetSample} lists the target sample, basic astrophysical parameters, and the date of the HST observations. The data were obtained with NIC1 in Multiaccum mode (NSAMP=18, STEP32), applying a two point dither pattern with the science targets centered in the top right or bottom right quadrant, respectively, and two telescope roll angles differing by 20 degree. The latter facilitates the application of the angular differential imaging technique for high-contrast subtraction of static PSF structures \citep{Mueller87,Lafreniere2007}. Total integration times in F110W and F160W amount to 1280\,s and 2560\,s, respectively, for each of the white dwarfs.

The raw data were pre-reduced with CALNICA Version 4.4.1. Compared to earlier reductions of this data set (\cite{Zinnecker06,Friedrich07,Zinnecker08}), the re-reduction included a temperature dependent dark correction. This resulted in a better background subtraction, and overall improved contrast ratios in particular at larger angular separations.

\begin{figure}
    \begin{center}
        \includegraphics[width=0.48\textwidth]{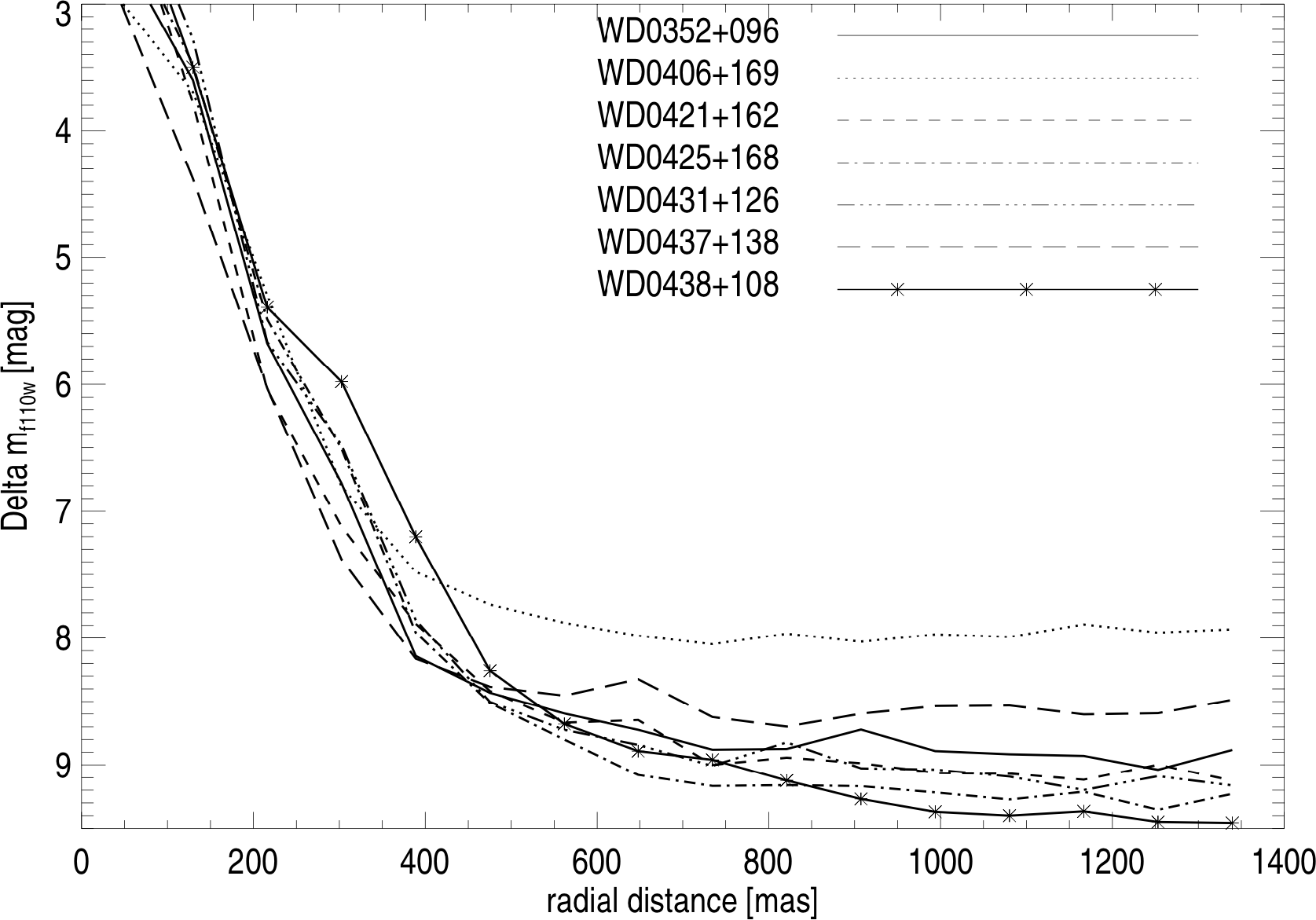}
    \end{center}
    \caption[]{Brightness contrast (3$\sigma$) in F110W vs.\ angular separation for the seven white dwarfs, reaching $\Delta {\rm m_{F110W}} > $ 8\,mag beyond 500\,mas).}
    \label{delta_f110w}
\end{figure}

The subsequent data reduction steps consisted of 
\begin{itemize}

\item[i]{pairwise subtraction of dithered positions to remove residual background, and visual inspection for wide companion candidates,}

\item[ii]{alignment and subtraction of individual 64 $\times$ 64 pixel subfields centered on the white dwarf for two roll angles to facilitate angular differential imaging (removal of static PSF pattern).}

\item[iii]{Combination and visual inspection of difference frames.  Real sources should show up as a positive and  negative PSF at the same separation from the white dwarf, and with instrumental position angles differing by 20 degree\footnote{For WD0421+162 the observations of the 1st HST orbit in F160W are all obtained at the same roll angle. We averaged all observations from this and the 2nd orbit with the same roll angle, and subtracted the one dithered data set obtained in the 2nd HST orbit at the 20 degree offset roll angle.} (see Figure~\ref{NIC1mosaic}).}

\item[iv]{Computation of the residual noise pattern in annuli centred on the white dwarf. The radial 3$\sigma$ brightest pixel detection limits for each white dwarf are shown in Figure~\ref{delta_f110w}.}
\end{itemize}

The contrast curves asymptotically approach the background limit for separations $\ge$0.6$''$. At a separation of 0.5$''$ the achievable contrast is still subject to PSF residuals. With the exception of WD0406+169, longer exposure times would not have significantly improved the contrast.
At angular distances $\ge$0.5$''$ from the white dwarf, the detection limit for the full sample is in the range m$_{\rm F110W}$ = 22.9 to 23.5\,mag and m$_{\rm F160W}$ = 22.2 to 22.9\,mag. 
The actual brightness differences (3$\sigma$ detection limits) at 0.5$''$ are $\Delta$m$_{\rm F110W}$ = 7.8 to 8.6\,mag, and $\Delta$m$_{\rm F160W}$ =  6.7 to 7.8\,mag (see Table \ref{DetectionLimits}).

Orbits of confirmed exoplanets with semi-major axis between 4 and 20\,A.U.\ have a mean eccentricity of $\approx$0.35.\footnote{\text{http://exoplanet.eu}} Assuming similar orbital parameters for our sample, we applied a correction factor of 1/0.83 for the conversion from projected separations to semimajor axis (see, e.g., \cite{Leinert1993} for a derivation).


\section{Flux and mass detection limits on exoplanets}

Starting from the 3$\sigma$ $\Delta$m$_{\rm F110W}$ and $\Delta$m$_{\rm F160W}$ contrast detection limits and the apparent 2MASS J and H magnitudes of the white dwarfs, we calculated the detection limits for companions in terms of apparent magnitudes. The apparent magnitudes were converted to absolute magnitudes using GAIA DR2 parallax measurements for comparison with theoretical evolutionary models for substellar objects.  

We selected evolutionary models for substellar objects with solar metallicity by  \cite{Baraffe03} and \cite{Spiegel12}. In the latter case, we consider ``hybrid'' (i.e. a linear combination of cloud and cloud-free) atmospheric models with either high-entropy (hothycl) or low-entropy (coldhycl) initial conditions.
For the transformation of brightness limits into mass limits the canonical age of the Hyades of 625\,Myr was assumed.
It is noted that according to the models even at this relatively advanced age, the most massive exoplanets still retain a ``memory'' of the starting conditions, i.e.\ the derived mass estimates assuming a cold start are typically 2\,M$_{\rm Jup}$ higher than for a hot start. The presence or absence of atmospheric condensation layers (``DUSTY'' or ``COND'' following the notation by \cite{Allard2001}), and variations of the metallicity result in uncertainties of the same order.
As the models by \cite{Spiegel12} are limited to masses $\le$10\,M$_{\rm Jup}$, in some cases we used an extrapolation to estimate mass limits. 

Also shown in Table \ref{DetectionLimits} are the mass limits derived from SPITZER IRAC observations in the 4.5\,$\mu$m band according to \cite{Farihi08}, which are based on the models by \cite{Baraffe03}, and assuming a uniform distance of 46\,pc towards all of the white dwarfs.

In all cases, observations in F110W provided lower mass limits, and thus tighter constraints than the observations in F160W. Overall mass limits are in the range of 4.6 to 6.7\,M$_{\rm Jup}$ assuming the models by \cite{Baraffe03}. Hybrid clouds and hot start models \citep{Spiegel12} place detection limits in the range 8 to 10\,M$_{\rm Jup}$. Thus irrespective of the model set we choose, we should be able to detect brown dwarf companions with masses $\ge$12\,M$_{\rm Jup}$ at current angular separations $\ge$0.5$''$ to any of the seven white dwarfs in the sample. In case the models by \cite{Baraffe03} are applicable, the average detection limit corresponds to exoplanets with masses in excess of $\approx$5.6\,M$_{\rm Jup}$, which represents an improved detection limit for resolved exoplanet companions compared to the (indirect) average limit of $\approx$9.4\,M$_{\rm Jup}$ from SPITZER/IRAC \citep{Farihi08}.

\begin{table*}
\caption{Detection limits at angular separation $\ge 0.5''$ in F110W and F160W: brightness difference, 3$\sigma$ apparent and absolute magnitude limit, corresponding upper mass limit for substellar companions according to evolutionary and atmospheric models, and current and initial projected physical separation corresponding to 0.5$''$ based on mass-loss estimates for the white dwarf progenitors and assuming a typical orbital eccentricity of 0.35. Mass limits at 4.5\,$\mu$m are from \citet{Farihi08} and are based on Baraffe models.}
\label{DetectionLimits}      
\centering                          
\begin{tabular}{l|r r r r r r r}        
name: WD...& 0352+096& 0406+169&0421+162&0425+168&0431+126&0437+138&0438+108\\ \hline
$\Delta$m$_{\rm F110W}$ [mag]&8.48 &7.79 &8.60 &8.59 &8.62 &8.35 &8.39 \\ 
m$_{\rm lim F110W}$ [mag]&23.31 &23.49 &23.35 &23.22 &23.39 &23.67 &22.89 \\ 
M$_{\rm lim F110W}$ [mag]&20.6 &20.0 &20.1 &19.8 &20.0 &20.4 &19.4 \\ 
mass$_{\rm Baraffe}$ [M$_{\rm Jupiter}$]&4.6 &5.7 &5.5 &6.2 &5.7 &4.9 & 6.7 \\ 
mass$_{\rm hot hycl}$ [M$_{\rm Jupiter}$]&7.7 &8.6 &8.5 &9.0 &8.6 &8.0 &9.6 \\ 
mass$_{\rm cold hycl}$ [M$_{\rm Jupiter}$]&9.1 &10.3: &10.1: &10.8: &10.3: &9.5 &11.7: \\ \hline
$\Delta$m$_{\rm F160W}$ [mag]&7.74 &6.70 &7.50 &7.80 &7.65 &7.35 &7.63 \\                                             
m$_{\rm lim F160W}$ [mag]&22.61 &22.18 &22.31 &22.45 &22.46 &22.86 &22.26 \\
M$_{\rm lim F160W}$ [mag]&19.9 &18.7 &19.0 &19.0 &19.1 &19.5 &18.8 \\
mass$_{\rm Baraffe}$ [M$_{\rm Jupiter}$]&5.9 &8.2 &7.2 &7.2 &7.0 &6.5 &7.8 \\ 
mass$_{\rm hot hycl}$ [M$_{\rm Jupiter}$]&8.3 &10.0 &9.5 &9.5 &9.4 &8.8 &9.8 \\ 
mass$_{\rm cold hycl}$ [M$_{\rm Jupiter}$]&9.8 &12.1: &11.5: &11.5: &11.3: &10.5: &11.9: \\ \hline
mass$_{4.5\mu m}$ [M$_{\rm Jupiter}$]&10 &7 &10 &10 &10 &8 &11 \\ \hline
a$_{\rm curr}$ [A.U.]& 21 & 30 & 27 & 29 & 28 & 28 & 30 \\
a$_{\rm init}$ [A.U.]& 4.7 & 7.4 & 6.5 & 7.3 & 6.9 & 6.0 & 7.8 \\
\hline                                   
\end{tabular}
     \begin{quote}
Notes: we selected models for solar metallicity by \cite{Baraffe03} and \cite{Spiegel12}. In the latter case, the mass estimates are based on models assuming hybrid clouds and either a hot (hothycl) or a cold (coldhycl) start.\\ ":" indicates that the mass limits are based on extrapolation beyond the mass range covered by the models.
      \end{quote}
\end{table*}


\section{Discussion}

\begin{figure}
    \begin{center}
        \includegraphics[width=0.48\textwidth]{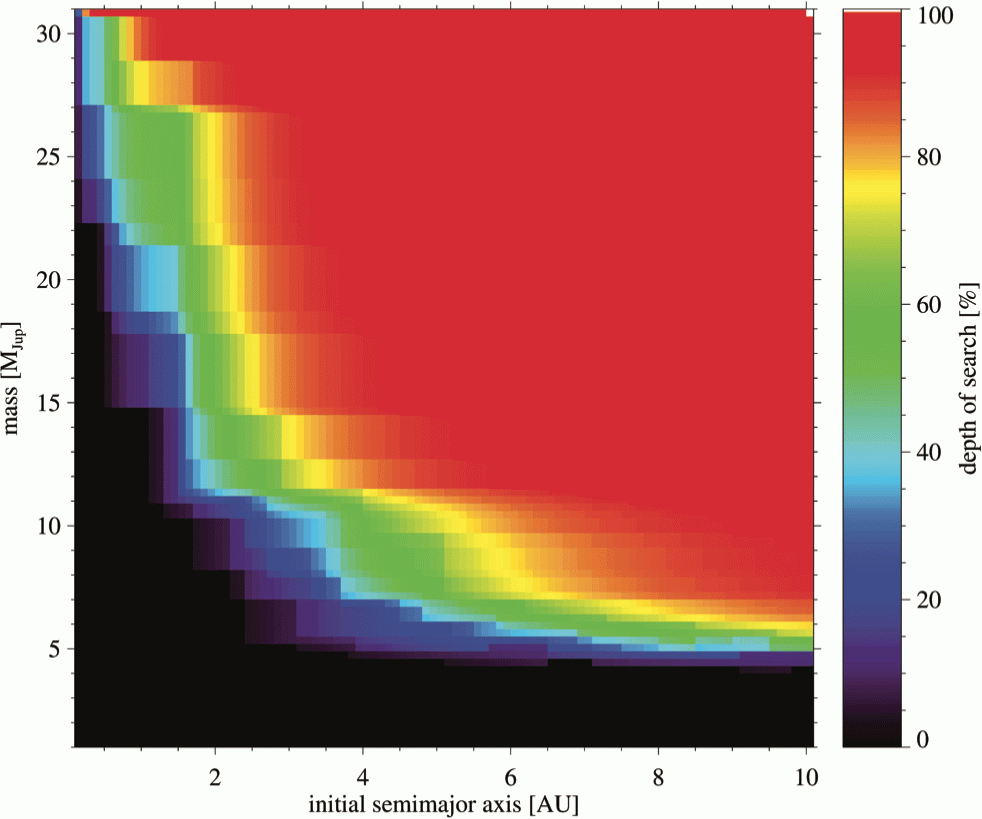}
    \end{center}
    \caption[]{Depth of search, indicating the completeness of the survey of the seven white dwarfs with respect to companion mass and semimajor axis for an assumed typical orbital eccentricity of 0.35. The green colour marks a detection probability of $\approx$50\%, while the red colour corresponds to $\gtrapprox$90\%.}
    \label{depthofsearch}
\end{figure}

Present-day mass estimates for Hyades white dwarfs have been derived from analysis of spectroscopic data using atmospheric models \citep{Bergeron2011,Gianninas2011}. We note that gravitational redshift measurements for a subset of the Hyades white dwarfs yield 5\% to 10\% lower masses estimates \citep{Pasquini2019}. For the following discussion, we consider this small systematic discrepancy in mass estimates of minor importance.

The Hyades white dwarf progenitors lost about 75\,\% of their initial mass \citep{Kalirai14}. Due to conservation of orbital angular momentum during the RG/AGB phase any surviving stellar or substellar companion must have migrated 
outward adiabatically, enlarging the semi-major axis of its orbit by a factor of 3.8 to 4.6 (ratio of initial stellar to final white dwarf mass, see Table \ref{TargetSample}). Giant planets with initial semi-major axis of their orbits $\gtrapprox$5\,A.U.\ would have survived the post-main sequence mass loss of the parent star (e.g., \cite{Villaver09,Nordhaus13}), and could now be found in orbits with semimajor axis $\gtrapprox$20 to 30\,A.U. The latter corresponds to angular separations $\ge$0.5$''$, which we probe in our HST/NICMOS survey. The present survey does not probe for planets initially in closer orbits ($<$2 to 3\,A.U.), which during the RG/AGB phase might have merged with the star due to dynamical friction and tidal interactions \citep{Mustill2013,Veras2014}.

The NICMOS observations confirm that all seven white dwarfs are single objects. Based on the detection limits and irrespective of evolutionary model we choose, we can rule out brown dwarf companions with masses $\ge$12\,M$_{\rm Jup}$ to any of the white dwarfs at current projected separations $\ge$18 to 25\,A.U.

Figure \ref{depthofsearch} visualizes the depth of search of our survey based on the models by \cite{Baraffe03}. We convert radial distances from the contrast curves (Figure \ref{delta_f110w}) into present day physical separations, and initial semimajor axis during the main-sequence phase of the host stars based on the white dwarf mass loss estimates.
The detection probabilities were calculated assuming planetary orbits with an eccentricity of 0.35 seen from 8000 uniformly distributed viewing angles and at 126 orbital phases distributed uniformly in time.
Compared to a search with SPITZER/IRAC using infrared excess of the white dwarfs as an indicator for unresolved very low-mass companions \citep{Farihi08}, our mass limits for resolved giant planets at angular separations $\ge$500\,mas are on average $\approx$4\,M$_{\rm Jup}$ lower.
Our survey reaches a 50\% depth of search at $\approx$6\,M$_{\rm Jup}$ and for an initial semimajor axis of 6\,AU. Ground-based adaptive optics assisted direct imaging surveys for giant exoplanets, which focus on young stars in the solar neigbourhood, in general probe an overlapping part of the parameter space.
For stars more massive than 1.5\,M$_\odot$, the GPIES survey at Gemini South \citep{Nielsen2019} reaches a 50\% depth of search at $\approx$6\,M$_{\rm Jup}$ for semimajor axis $\ge$50\,AU.
The SHINE survey at the VLT \citep{Vigan2020} reports for spectral types M to B a 50\% depth of survey at 6 M$_{\rm Jup}$ for semimajor axis $\ge$15 to 20\,AU.

While our sample of seven single white dwarfs is small, it provides further evidence that initially dense cluster environments, which included O and early B stars,  might not be highly conducive for the formation and transformation of massive circumstellar disks around stars with masses $\ge$2.8 M$_\odot$ into giant planets with semimajor axes $\ge$6 A.U and with $\ge 6$ Jupiter masses. This is in agreement with the radial velocity survey for exoplanets around 373 G- and K-type giants by \cite{Reffert15}, which did not find any planets around giants with initial main sequence masses higher than 2.7\,M$_\odot$. Other high contrast direct imaging searches for planetary mass companions to nearby white dwarfs also did not yield any direct detections \citep{Debes05,Farihi05,Debes06,Xu2015} apart from GD165AB.

{\it Indirect} evidence for the presence of planets has been derived from abundance anomalies in white dwarf atmospheres and in their accretion disks as well as from the longevity of debris disks (e.g.\ \cite{Zuckerman2010,Zuckerman13,Farihi2013,Bergfors2014,Xu2019}).
Using  optical spectroscopy, \cite{Gaensicke2019} reported strong evidence for a photoevaporating giant planet in close (15 R$_\odot$) orbit around a hot white dwarf (WD 1145+017); see also an earlier case suggested by \cite{Chu2001}. Recently \cite{Vanderburg2020} announced the detection of a transiting giant planet in a 1.4 day orbit around the relatively old WD 1856+534.

In the Hyades, several Earth to Neptune sized exoplanets and exoplanet candidates transiting K- and M-dwarfs were identified with K2 \citep{Mann16,Ciardi18,Livingstone18,Vanderburg18}. The 2.7\,M$_\odot$ K-giant $\epsilon$\,Tau still remains the most massive Hyades member known to host a (giant) exoplanet \citep{Sato07}.

The divergence of mass estimates between the different evolutionary models and the underlying uncertainty about the starting conditions in the formation of substellar objects highlights the importance of testing and calibrating evolutionary models. This requires both independent mass estimates via astrometric or radial velocity reflex motion of the host star, and studies of the early formation phases of exoplanets.
ELTs equipped with laser guide star adaptive optics systems could probe the Hyades white dwarfs for giant planets at closer projected separations, and lower mass limits. ELT/MICADO is expected to achieve a contrast of $\Delta {\rm H} \approx$14.8\,mag at an angular separation of 500\,mas \citep{Perrot2018}, and thus (according to the model by \cite{Baraffe03}) could detect planets with 1 to 2\,M$_{\rm Jup}$ orbiting a white dwarf in the Hyades cluster . In the K-band JWST/NIRCam is predicted to achieve a contrast of $\Delta {\rm m}_{\rm F210M} \approx 11.5$\,mag at 500\,mas \citep{Beichman2010}, which yields mass detection limits of the same order as the present HST/NICMOS study in F110W (giant planets in the mass and age range under consideration are considerably fainter in the K-band than in the J- and H-bands due to molecular opacities in their atmospheres). The larger collecting area, and improved detector technology of JWST/MIRI relative to SPITZER/IRAC will provide a considerably improved sensitivity for detecting infrared excess from unresolved companions, and might also reach detection limits of 1 to 2\,M$_{\rm Jup}$ for the Hyades white dwarfs.


\section*{Acknowledgements}

We thank S.\ Reffert, S.\ R\"oser, J.\ van Cleve, and S.\ Xu for helpful comments on a draft of the paper.

Based on observations made with the NASA/ESA Hubble Space Telescope (GO 9737), obtained from the data archive at the Space Telescope Science Institute. STScI is operated by the Association of Universities for Research in Astronomy, Inc. under NASA contract NAS 5-26555.

This publication makes use of data products from the Two Micron All Sky Survey, which is a joint project of the University of Massachusetts and the Infrared Processing and Analysis Center/California Institute of Technology, funded by the National Aeronautics and Space Administration and the National Science Foundation.

This work has made use of data from the European Space Agency (ESA) mission
{\it Gaia} (\url{https://www.cosmos.esa.int/gaia}), processed by the {\it Gaia}
Data Processing and Analysis Consortium (DPAC,
\url{https://www.cosmos.esa.int/web/gaia/dpac/consortium}). Funding for the DPAC
has been provided by national institutions, in particular the institutions
participating in the {\it Gaia} Multilateral Agreement.

\section*{Data availability}

The data underlying this article is available online from the Barbara A. Mikulski Archive for Space Telescopes. The final data products will be shared by the corresponding author on request.



\bibliographystyle{mnras}
\bibliography{lit} 

\begin{thebibliography}{}
\makeatletter
\relax
\def\mn@urlcharsother{\let\do\@makeother \do\$\do\&\do\#\do\^\do\_\do\%\do\~}
\def\mn@doi{\begingroup\mn@urlcharsother \@ifnextchar [ {\mn@doi@}
  {\mn@doi@[]}}
\def\mn@doi@[#1]#2{\def\@tempa{#1}\ifx\@tempa\@empty \href
  {http://dx.doi.org/#2} {doi:#2}\else \href {http://dx.doi.org/#2} {#1}\fi
  \endgroup}
\def\mn@eprint#1#2{\mn@eprint@#1:#2::\@nil}
\def\mn@eprint@arXiv#1{\href {http://arxiv.org/abs/#1} {{\tt arXiv:#1}}}
\def\mn@eprint@dblp#1{\href {http://dblp.uni-trier.de/rec/bibtex/#1.xml}
  {dblp:#1}}
\def\mn@eprint@#1:#2:#3:#4\@nil{\def\@tempa {#1}\def\@tempb {#2}\def\@tempc
  {#3}\ifx \@tempc \@empty \let \@tempc \@tempb \let \@tempb \@tempa \fi \ifx
  \@tempb \@empty \def\@tempb {arXiv}\fi \@ifundefined
  {mn@eprint@\@tempb}{\@tempb:\@tempc}{\expandafter \expandafter \csname
  mn@eprint@\@tempb\endcsname \expandafter{\@tempc}}}

\bibitem[\protect\citeauthoryear{{Allard}, {Hauschildt}, {Alexander}, {Tamanai}
   \& {Schweitzer}}{{Allard} et~al.}{2001}]{Allard2001}
{Allard} F.,  {Hauschildt} P.~H.,  {Alexander} D.~R.,  {Tamanai} A.,
  {Schweitzer} A.,  2001, \mn@doi [\apj] {10.1086/321547}, \href
  {https://ui.adsabs.harvard.edu/abs/2001ApJ...556..357A} {556, 357}

\bibitem[\protect\citeauthoryear{{Bailer-Jones}, {Rybizki}, {Fouesneau},
  {Mantelet}  \& {Andrae}}{{Bailer-Jones} et~al.}{2018}]{BailerJones18}
{Bailer-Jones} C.~A.~L.,  {Rybizki} J.,  {Fouesneau} M.,  {Mantelet} G.,
  {Andrae} R.,  2018, \mn@doi [\aj] {10.3847/1538-3881/aacb21}, \href
  {http://adsabs.harvard.edu/abs/2018AJ....156...58B} {156, 58}

\bibitem[\protect\citeauthoryear{{Baraffe}, {Chabrier}, {Barman}, {Allard}  \&
  {Hauschildt}}{{Baraffe} et~al.}{2003}]{Baraffe03}
{Baraffe} I.,  {Chabrier} G.,  {Barman} T.~S.,  {Allard} F.,   {Hauschildt}
  P.~H.,  2003, \mn@doi [\aap] {10.1051/0004-6361:20030252}, \href
  {http://adsabs.harvard.edu/abs/2003A%26A...402..701B} {402, 701}

\bibitem[\protect\citeauthoryear{{Becklin} \& {Zuckerman}}{{Becklin} \&
  {Zuckerman}}{1988}]{Becklin88}
{Becklin} E.~E.,  {Zuckerman} B.,  1988, \mn@doi [\nat] {10.1038/336656a0},
  \href {http://esoads.eso.org/abs/1988Natur.336..656B} {336, 656}

\bibitem[\protect\citeauthoryear{{Beichman} et~al.,}{{Beichman}
  et~al.}{2010}]{Beichman2010}
{Beichman} C.~A.,  et~al., 2010, \mn@doi [\pasp] {10.1086/651057}, \href
  {https://ui.adsabs.harvard.edu/abs/2010PASP..122..162B} {122, 162}

\bibitem[\protect\citeauthoryear{{Bergeron} et~al.,}{{Bergeron}
  et~al.}{2011}]{Bergeron2011}
{Bergeron} P.,  et~al., 2011, \mn@doi [\apj] {10.1088/0004-637X/737/1/28},
  \href {https://ui.adsabs.harvard.edu/abs/2011ApJ...737...28B} {737, 28}

\bibitem[\protect\citeauthoryear{{Bergfors}, {Farihi}, {Dufour}  \&
  {Rocchetto}}{{Bergfors} et~al.}{2014}]{Bergfors2014}
{Bergfors} C.,  {Farihi} J.,  {Dufour} P.,   {Rocchetto} M.,  2014, \mn@doi
  [\mnras] {10.1093/mnras/stu1565}, \href
  {https://ui.adsabs.harvard.edu/abs/2014MNRAS.444.2147B} {444, 2147}

\bibitem[\protect\citeauthoryear{{Burleigh}, {Clarke}  \& {Hodgkin}}{{Burleigh}
  et~al.}{2002}]{Burleigh02}
{Burleigh} M.~R.,  {Clarke} F.~J.,   {Hodgkin} S.~T.,  2002, \mn@doi [\mnras]
  {10.1046/j.1365-8711.2002.05417.x}, \href
  {http://adsabs.harvard.edu/abs/2002MNRAS.331L..41B} {331, L41}

\bibitem[\protect\citeauthoryear{{Chauvin} et~al.,}{{Chauvin}
  et~al.}{2005}]{Chauvin05}
{Chauvin} G.,  et~al., 2005, \mn@doi [A\&A] {10.1051/0004-6361:200500111},
  \href {http://cdsads.u-strasbg.fr/abs/2005A%26A...438L..29C} {438, L29}

\bibitem[\protect\citeauthoryear{{Chauvin} et~al.,}{{Chauvin}
  et~al.}{2017}]{Chauvin17}
{Chauvin} G.,  et~al., 2017, \mn@doi [\aap] {10.1051/0004-6361/201731152},
  \href {http://adsabs.harvard.edu/abs/2017A%26A...605L...9C} {605, L9}

\bibitem[\protect\citeauthoryear{{Chu}, {Dunne}, {Gruendl}  \&
  {Brandner}}{{Chu} et~al.}{2001}]{Chu2001}
{Chu} Y.-H.,  {Dunne} B.~C.,  {Gruendl} R.~A.,   {Brandner} W.,  2001, \mn@doi
  [\apjl] {10.1086/318058}, \href
  {https://ui.adsabs.harvard.edu/abs/2001ApJ...546L..61C} {546, L61}

\bibitem[\protect\citeauthoryear{{Ciardi} et~al.,}{{Ciardi}
  et~al.}{2018}]{Ciardi18}
{Ciardi} D.~R.,  et~al., 2018, \mn@doi [\aj] {10.3847/1538-3881/aa9921}, \href
  {http://esoads.eso.org/abs/2018AJ....155...10C} {155, 10}

\bibitem[\protect\citeauthoryear{{Debes}, {Sigurdsson}  \& {Woodgate}}{{Debes}
  et~al.}{2005}]{Debes05}
{Debes} J.~H.,  {Sigurdsson} S.,   {Woodgate} B.~E.,  2005, \mn@doi [\aj]
  {10.1086/432660}, \href {http://adsabs.harvard.edu/abs/2005AJ....130.1221D}
  {130, 1221}

\bibitem[\protect\citeauthoryear{{Debes}, {Ge}  \& {Ftaclas}}{{Debes}
  et~al.}{2006}]{Debes06}
{Debes} J.~H.,  {Ge} J.,   {Ftaclas} C.,  2006, \mn@doi [\aj] {10.1086/498740},
  \href {http://adsabs.harvard.edu/abs/2006AJ....131..640D} {131, 640}

\bibitem[\protect\citeauthoryear{{Farihi}, {Zuckerman}  \& {Becklin}}{{Farihi}
  et~al.}{2005}]{Farihi05}
{Farihi} J.,  {Zuckerman} B.,   {Becklin} E.~E.,  2005, \mn@doi [Astronomische
  Nachrichten] {10.1002/asna.200510434}, \href
  {http://adsabs.harvard.edu/abs/2005AN....326..964F} {326, 964}

\bibitem[\protect\citeauthoryear{{Farihi}, {Becklin}  \& {Zuckerman}}{{Farihi}
  et~al.}{2008}]{Farihi08}
{Farihi} J.,  {Becklin} E.~E.,   {Zuckerman} B.,  2008, \mn@doi [\apj]
  {10.1086/588726}, \href {http://adsabs.harvard.edu/abs/2008ApJ...681.1470F}
  {681, 1470}

\bibitem[\protect\citeauthoryear{{Farihi}, {G{\"a}nsicke}  \&
  {Koester}}{{Farihi} et~al.}{2013}]{Farihi2013}
{Farihi} J.,  {G{\"a}nsicke} B.~T.,   {Koester} D.,  2013, \mn@doi [\mnras]
  {10.1093/mnras/stt432}, \href
  {https://ui.adsabs.harvard.edu/abs/2013MNRAS.432.1955F} {432, 1955}

\bibitem[\protect\citeauthoryear{{Friedrich}, {Zinnecker}, {Correia},
  {Brandner}, {Burleigh}  \& {McCaughrean}}{{Friedrich}
  et~al.}{2007}]{Friedrich07}
{Friedrich} S.,  {Zinnecker} H.,  {Correia} S.,  {Brandner} W.,  {Burleigh} M.,
    {McCaughrean} M.,  2007, in {Napiwotzki} R.,  {Burleigh} M.~R.,  eds,
  Astronomical Society of the Pacific Conference Series Vol. 372, 15th European
  Workshop on White Dwarfs. p.~343 (\mn@eprint {} {astro-ph/0611511})

\bibitem[\protect\citeauthoryear{{Gaia Collaboration} et~al.,}{{Gaia
  Collaboration} et~al.}{2016}]{GAIA2016}
{Gaia Collaboration} et~al., 2016, \mn@doi [\aap]
  {10.1051/0004-6361/201629272}, \href
  {http://adsabs.harvard.edu/abs/2016A%26A...595A...1G} {595, A1}

\bibitem[\protect\citeauthoryear{{Gaia Collaboration} et~al.,}{{Gaia
  Collaboration} et~al.}{2018}]{Brown18}
{Gaia Collaboration} et~al., 2018, \mn@doi [\aap]
  {10.1051/0004-6361/201833051}, \href
  {https://ui.adsabs.harvard.edu/abs/2018A&A...616A...1G} {616, A1}

\bibitem[\protect\citeauthoryear{{G{\"a}nsicke}, {Schreiber}, {Toloza},
  {Fusillo}, {Koester}  \& {Manser}}{{G{\"a}nsicke}
  et~al.}{2019}]{Gaensicke2019}
{G{\"a}nsicke} B.~T.,  {Schreiber} M.~R.,  {Toloza} O.,  {Fusillo} N. P.~G.,
  {Koester} D.,   {Manser} C.~J.,  2019, \mn@doi [\nat]
  {10.1038/s41586-019-1789-8}, \href
  {https://ui.adsabs.harvard.edu/abs/2019Natur.576...61G} {576, 61}

\bibitem[\protect\citeauthoryear{{Gianninas}, {Bergeron}  \&
  {Ruiz}}{{Gianninas} et~al.}{2011}]{Gianninas2011}
{Gianninas} A.,  {Bergeron} P.,   {Ruiz} M.~T.,  2011, \mn@doi [\apj]
  {10.1088/0004-637X/743/2/138}, \href
  {https://ui.adsabs.harvard.edu/abs/2011ApJ...743..138G} {743, 138}

\bibitem[\protect\citeauthoryear{{Gilliland} et~al.,}{{Gilliland}
  et~al.}{2000}]{Gilliland00}
{Gilliland} R.~L.,  et~al., 2000, \mn@doi [ApJL] {10.1086/317334}, \href
  {http://adsabs.harvard.edu/abs/2000ApJ...545L..47G} {545, L47}

\bibitem[\protect\citeauthoryear{{Gould} \& {Kilic}}{{Gould} \&
  {Kilic}}{2008}]{Gould08}
{Gould} A.,  {Kilic} M.,  2008, \mn@doi [\apjl] {10.1086/527476}, \href
  {http://adsabs.harvard.edu/abs/2008ApJ...673L..75G} {673, L75}

\bibitem[\protect\citeauthoryear{{Guenther}, {Paulson}, {Cochran}, {Patience},
  {Hatzes}  \& {Macintosh}}{{Guenther} et~al.}{2005}]{Guenther05}
{Guenther} E.~W.,  {Paulson} D.~B.,  {Cochran} W.~D.,  {Patience} J.,  {Hatzes}
  A.~P.,   {Macintosh} B.,  2005, \mn@doi [\aap] {10.1051/0004-6361:20052851},
  \href {http://esoads.eso.org/abs/2005A%26A...442.1031G} {442, 1031}

\bibitem[\protect\citeauthoryear{{Hogan}, {Burleigh}  \& {Clarke}}{{Hogan}
  et~al.}{2011}]{Hogan2011}
{Hogan} E.,  {Burleigh} M.~R.,   {Clarke} F.~J.,  2011, in {Schuh} S.,
  {Drechsel} H.,   {Heber} U.,  eds,  American Institute of Physics Conference
  Series Vol. 1331, American Institute of Physics Conference Series. pp
  271--277 (\mn@eprint {arXiv} {1102.0506}), \mn@doi{10.1063/1.3556210}

\bibitem[\protect\citeauthoryear{{Kalirai}, {Marigo}  \& {Tremblay}}{{Kalirai}
  et~al.}{2014}]{Kalirai14}
{Kalirai} J.~S.,  {Marigo} P.,   {Tremblay} P.-E.,  2014, \mn@doi [\apj]
  {10.1088/0004-637X/782/1/17}, \href
  {http://adsabs.harvard.edu/abs/2014ApJ...782...17K} {782, 17}

\bibitem[\protect\citeauthoryear{{Kennedy} \& {Kenyon}}{{Kennedy} \&
  {Kenyon}}{2008}]{Kennedy2008}
{Kennedy} G.~M.,  {Kenyon} S.~J.,  2008, \mn@doi [\apj] {10.1086/524130}, \href
  {https://ui.adsabs.harvard.edu/abs/2008ApJ...673..502K} {673, 502}

\bibitem[\protect\citeauthoryear{{Keppler} et~al.,}{{Keppler}
  et~al.}{2018}]{Keppler18}
{Keppler} M.,  et~al., 2018, \mn@doi [\aap] {10.1051/0004-6361/201832957},
  \href {https://ui.adsabs.harvard.edu/abs/2018A&A...617A..44K} {617, A44}

\bibitem[\protect\citeauthoryear{{Kopytova}, {Brandner}, {Tognelli}, {Prada
  Moroni}, {Da Rio}, {R{\"o}ser}  \& {Schilbach}}{{Kopytova}
  et~al.}{2016}]{Kopytova16}
{Kopytova} T.~G.,  {Brandner} W.,  {Tognelli} E.,  {Prada Moroni} P.~G.,  {Da
  Rio} N.,  {R{\"o}ser} S.,   {Schilbach} E.,  2016, \mn@doi [\aap]
  {10.1051/0004-6361/201527044}, \href
  {http://esoads.eso.org/abs/2016A%26A...585A...7K} {585, A7}

\bibitem[\protect\citeauthoryear{{Kov{\'a}cs} et~al.,}{{Kov{\'a}cs}
  et~al.}{2014}]{Kovacs14}
{Kov{\'a}cs} G.,  et~al., 2014, \mn@doi [\mnras] {10.1093/mnras/stu946}, \href
  {http://adsabs.harvard.edu/abs/2014MNRAS.442.2081K} {442, 2081}

\bibitem[\protect\citeauthoryear{{Krist}, {Hook}  \& {Stoehr}}{{Krist}
  et~al.}{2011}]{Krist2011}
{Krist} J.~E.,  {Hook} R.~N.,   {Stoehr} F.,  2011, in Optical Modeling and
  Performance Predictions V. p. 81270J, \mn@doi{10.1117/12.892762}

\bibitem[\protect\citeauthoryear{{Lafreni{\`e}re}, {Marois}, {Doyon}, {Nadeau}
  \& {Artigau}}{{Lafreni{\`e}re} et~al.}{2007}]{Lafreniere2007}
{Lafreni{\`e}re} D.,  {Marois} C.,  {Doyon} R.,  {Nadeau} D.,   {Artigau}
  {\'E}.,  2007, \mn@doi [\apj] {10.1086/513180}, \href
  {https://ui.adsabs.harvard.edu/abs/2007ApJ...660..770L} {660, 770}

\bibitem[\protect\citeauthoryear{{Lagrange} et~al.,}{{Lagrange}
  et~al.}{2009}]{Lagrange09}
{Lagrange} A.-M.,  et~al., 2009, \mn@doi [A\&A] {10.1051/0004-6361:200811325},
  \href {http://adsabs.harvard.edu/abs/2009A%26A...493L..21L} {493, L21}

\bibitem[\protect\citeauthoryear{{Lagrange} et~al.,}{{Lagrange}
  et~al.}{2010}]{Lagrange2010}
{Lagrange} A.~M.,  et~al., 2010, \mn@doi [Science] {10.1126/science.1187187},
  \href {https://ui.adsabs.harvard.edu/abs/2010Sci...329...57L} {329, 57}

\bibitem[\protect\citeauthoryear{{Leinert}, {Zinnecker}, {Weitzel}, {Christou},
  {Ridgway}, {Jameson}, {Haas}  \& {Lenzen}}{{Leinert}
  et~al.}{1993}]{Leinert1993}
{Leinert} C.,  {Zinnecker} H.,  {Weitzel} N.,  {Christou} J.,  {Ridgway} S.~T.,
   {Jameson} R.,  {Haas} M.,   {Lenzen} R.,  1993, \aap, \href
  {https://ui.adsabs.harvard.edu/abs/1993A&A...278..129L} {278, 129}

\bibitem[\protect\citeauthoryear{{Livingston} et~al.,}{{Livingston}
  et~al.}{2018}]{Livingstone18}
{Livingston} J.~H.,  et~al., 2018, \mn@doi [\aj] {10.3847/1538-3881/aaa841},
  \href {http://esoads.eso.org/abs/2018AJ....155..115L} {155, 115}

\bibitem[\protect\citeauthoryear{{Luhman}, {Burgasser}  \&
  {Bochanski}}{{Luhman} et~al.}{2011}]{Luhman2011}
{Luhman} K.~L.,  {Burgasser} A.~J.,   {Bochanski} J.~J.,  2011, \mn@doi [\apjl]
  {10.1088/2041-8205/730/1/L9}, \href
  {https://ui.adsabs.harvard.edu/abs/2011ApJ...730L...9L} {730, L9}

\bibitem[\protect\citeauthoryear{{Macintosh} et~al.,}{{Macintosh}
  et~al.}{2015}]{Macintosh15}
{Macintosh} B.,  et~al., 2015, \mn@doi [Science] {10.1126/science.aac5891},
  \href {http://adsabs.harvard.edu/abs/2015Sci...350...64M} {350, 64}

\bibitem[\protect\citeauthoryear{{Mann} et~al.,}{{Mann} et~al.}{2016}]{Mann16}
{Mann} A.~W.,  et~al., 2016, \mn@doi [\apj] {10.3847/0004-637X/818/1/46}, \href
  {http://esoads.eso.org/abs/2016ApJ...818...46M} {818, 46}

\bibitem[\protect\citeauthoryear{{Marois}, {Macintosh}, {Barman}, {Zuckerman},
  {Song}, {Patience}, {Lafreni{\`e}re}  \& {Doyon}}{{Marois}
  et~al.}{2008}]{Marois08}
{Marois} C.,  {Macintosh} B.,  {Barman} T.,  {Zuckerman} B.,  {Song} I.,
  {Patience} J.,  {Lafreni{\`e}re} D.,   {Doyon} R.,  2008, \mn@doi [Science]
  {10.1126/science.1166585}, \href
  {http://adsabs.harvard.edu/abs/2008Sci...322.1348M} {322, 1348}

\bibitem[\protect\citeauthoryear{{Marois}, {Zuckerman}, {Konopacky},
  {Macintosh}  \& {Barman}}{{Marois} et~al.}{2010}]{Marois2010}
{Marois} C.,  {Zuckerman} B.,  {Konopacky} Q.~M.,  {Macintosh} B.,   {Barman}
  T.,  2010, \mn@doi [\nat] {10.1038/nature09684}, \href
  {https://ui.adsabs.harvard.edu/abs/2010Natur.468.1080M} {468, 1080}

\bibitem[\protect\citeauthoryear{{M\"uller} \& {Weigelt}}{{M\"uller} \&
  {Weigelt}}{1987}]{Mueller87}
{M\"uller} M.,  {Weigelt} G.,  1987, \aap, \href
  {http://adsabs.harvard.edu/abs/1987A%26A...175..312M} {175, 312}

\bibitem[\protect\citeauthoryear{{Mustill}, {Villaver}, {Veras}, {Bonsor}  \&
  {Wyatt}}{{Mustill} et~al.}{2013}]{Mustill2013}
{Mustill} A.~J.,  {Villaver} E.,  {Veras} D.,  {Bonsor} A.,   {Wyatt} M.~C.,
  2013, in European Physical Journal Web of Conferences. p. 06008,
  \mn@doi{10.1051/epjconf/20134706008}

\bibitem[\protect\citeauthoryear{{Nielsen} et~al.,}{{Nielsen}
  et~al.}{2019}]{Nielsen2019}
{Nielsen} E.~L.,  et~al., 2019, \mn@doi [\aj] {10.3847/1538-3881/ab16e9}, \href
  {https://ui.adsabs.harvard.edu/abs/2019AJ....158...13N} {158, 13}

\bibitem[\protect\citeauthoryear{{Nordhaus} \& {Spiegel}}{{Nordhaus} \&
  {Spiegel}}{2013}]{Nordhaus13}
{Nordhaus} J.,  {Spiegel} D.~S.,  2013, \mn@doi [\mnras]
  {10.1093/mnras/stt569}, \href
  {http://adsabs.harvard.edu/abs/2013MNRAS.432..500N} {432, 500}

\bibitem[\protect\citeauthoryear{{Pasquini}, {Pala}, {Ludwig}, {Leao}, {de
  Medeiros}  \& {Weiss}}{{Pasquini} et~al.}{2019}]{Pasquini2019}
{Pasquini} L.,  {Pala} A.~F.,  {Ludwig} H.~G.,  {Leao} I.~C.,  {de Medeiros}
  J.~R.,   {Weiss} A.,  2019, \mn@doi [\aap] {10.1051/0004-6361/201935835},
  \href {https://ui.adsabs.harvard.edu/abs/2019A&A...627L...8P} {627, L8}

\bibitem[\protect\citeauthoryear{{Paulson}, {Saar}, {Cochran}  \&
  {Henry}}{{Paulson} et~al.}{2004}]{Paulson04}
{Paulson} D.~B.,  {Saar} S.~H.,  {Cochran} W.~D.,   {Henry} G.~W.,  2004,
  \mn@doi [\aj] {10.1086/381948}, \href
  {http://esoads.eso.org/abs/2004AJ....127.1644P} {127, 1644}

\bibitem[\protect\citeauthoryear{{Perrot}, {Baudoz}, {Boccaletti}, {Rousset},
  {Huby}, {Cl{\'e}net}, {Durand }  \& {Davies}}{{Perrot}
  et~al.}{2018}]{Perrot2018}
{Perrot} C.,  {Baudoz} P.,  {Boccaletti} A.,  {Rousset} G.,  {Huby} E.,
  {Cl{\'e}net} Y.,  {Durand } S.,   {Davies} R.,  2018, \mn@doi [arXiv
  e-prints] {10.26698/AO4ELT5.0159}, \href
  {https://ui.adsabs.harvard.edu/abs/2018arXiv180401371P} {p. arXiv:1804.01371}

\bibitem[\protect\citeauthoryear{{Perryman}}{{Perryman}}{2018}]{Perryman2018}
{Perryman} M.,  2018, {The Exoplanet Handbook}

\bibitem[\protect\citeauthoryear{{Perryman} et~al.,}{{Perryman}
  et~al.}{1998}]{Perryman98}
{Perryman} M.~A.~C.,  et~al., 1998, A\&A, \href
  {http://adsabs.harvard.edu/abs/1998A%26A...331...81P} {331, 81}

\bibitem[\protect\citeauthoryear{{Reffert}, {Bergmann}, {Quirrenbach},
  {Trifonov}  \& {K{\"u}nstler}}{{Reffert} et~al.}{2015}]{Reffert15}
{Reffert} S.,  {Bergmann} C.,  {Quirrenbach} A.,  {Trifonov} T.,
  {K{\"u}nstler} A.,  2015, \mn@doi [\aap] {10.1051/0004-6361/201322360}, \href
  {http://adsabs.harvard.edu/abs/2015A%26A...574A.116R} {574, A116}

\bibitem[\protect\citeauthoryear{{R{\"o}ser}, {Schilbach}, {Piskunov},
  {Kharchenko}  \& {Scholz}}{{R{\"o}ser} et~al.}{2011}]{Roeser11}
{R{\"o}ser} S.,  {Schilbach} E.,  {Piskunov} A.~E.,  {Kharchenko} N.~V.,
  {Scholz} R.-D.,  2011, \mn@doi [A\&A] {10.1051/0004-6361/201116948}, \href
  {http://adsabs.harvard.edu/abs/2011A%26A...531A..92R} {531, A92}

\bibitem[\protect\citeauthoryear{{Salaris} \& {Bedin}}{{Salaris} \&
  {Bedin}}{2018}]{Salaris2018}
{Salaris} M.,  {Bedin} L.~R.,  2018, \mn@doi [\mnras] {10.1093/mnras/sty2096},
  \href {https://ui.adsabs.harvard.edu/abs/2018MNRAS.480.3170S} {480, 3170}

\bibitem[\protect\citeauthoryear{{Sato} et~al.,}{{Sato} et~al.}{2007}]{Sato07}
{Sato} B.,  et~al., 2007, \mn@doi [\apj] {10.1086/513503}, \href
  {http://esoads.eso.org/abs/2007ApJ...661..527S} {661, 527}

\bibitem[\protect\citeauthoryear{{Schilbach} \& {R{\"o}ser}}{{Schilbach} \&
  {R{\"o}ser}}{2012}]{Schilbach12}
{Schilbach} E.,  {R{\"o}ser} S.,  2012, \mn@doi [\aap]
  {10.1051/0004-6361/201117688}, \href
  {http://adsabs.harvard.edu/abs/2012A%26A...537A.129S} {537, A129}

\bibitem[\protect\citeauthoryear{{Shabram} et~al.,}{{Shabram}
  et~al.}{2020}]{Shabram2020}
{Shabram} M.~I.,  et~al., 2020, \mn@doi [\aj] {10.3847/1538-3881/ab90fe}, \href
  {https://ui.adsabs.harvard.edu/abs/2020AJ....160...16S} {160, 16}

\bibitem[\protect\citeauthoryear{{Spiegel} \& {Burrows}}{{Spiegel} \&
  {Burrows}}{2012}]{Spiegel12}
{Spiegel} D.~S.,  {Burrows} A.,  2012, \mn@doi [\apj]
  {10.1088/0004-637X/745/2/174}, \href
  {http://adsabs.harvard.edu/abs/2012ApJ...745..174S} {745, 174}

\bibitem[\protect\citeauthoryear{{Tremblay}, {Schilbach}, {R{\"o}ser},
  {Jordan}, {Ludwig}  \& {Goldman}}{{Tremblay} et~al.}{2012}]{Tremblay12}
{Tremblay} P.-E.,  {Schilbach} E.,  {R{\"o}ser} S.,  {Jordan} S.,  {Ludwig}
  H.-G.,   {Goldman} B.,  2012, \mn@doi [A\&A] {10.1051/0004-6361/201220057},
  \href {http://adsabs.harvard.edu/abs/2012A%26A...547A..99T} {547, A99}

\bibitem[\protect\citeauthoryear{{Vanderburg} et~al.,}{{Vanderburg}
  et~al.}{2018}]{Vanderburg18}
{Vanderburg} A.,  et~al., 2018, \mn@doi [\aj] {10.3847/1538-3881/aac894}, \href
  {http://esoads.eso.org/abs/2018AJ....156...46V} {156, 46}

\bibitem[\protect\citeauthoryear{{Vanderburg} et~al.,}{{Vanderburg}
  et~al.}{2020}]{Vanderburg2020}
{Vanderburg} A.,  et~al., 2020, arXiv e-prints, \href
  {https://ui.adsabs.harvard.edu/abs/2020arXiv200907282V} {p. arXiv:2009.07282}

\bibitem[\protect\citeauthoryear{{Veras}, {Evans}, {Wyatt}  \& {Tout}}{{Veras}
  et~al.}{2014}]{Veras2014}
{Veras} D.,  {Evans} N.~W.,  {Wyatt} M.~C.,   {Tout} C.~A.,  2014, \mn@doi
  [\mnras] {10.1093/mnras/stt1905}, \href
  {https://ui.adsabs.harvard.edu/abs/2014MNRAS.437.1127V} {437, 1127}

\bibitem[\protect\citeauthoryear{{Vigan} et~al.,}{{Vigan}
  et~al.}{2020}]{Vigan2020}
{Vigan} A.,  et~al., 2020, arXiv e-prints, \href
  {https://ui.adsabs.harvard.edu/abs/2020arXiv200706573V} {p. arXiv:2007.06573}

\bibitem[\protect\citeauthoryear{{Villaver} \& {Livio}}{{Villaver} \&
  {Livio}}{2009}]{Villaver09}
{Villaver} E.,  {Livio} M.,  2009, \mn@doi [\apjl]
  {10.1088/0004-637X/705/1/L81}, \href
  {http://esoads.eso.org/abs/2009ApJ...705L..81V} {705, L81}

\bibitem[\protect\citeauthoryear{{Xu}, {Ertel}, {Wahhaj}, {Milli}, {Scicluna}
  \& {Bertrang}}{{Xu} et~al.}{2015}]{Xu2015}
{Xu} S.,  {Ertel} S.,  {Wahhaj} Z.,  {Milli} J.,  {Scicluna} P.,   {Bertrang}
  G.~H.~M.,  2015, \mn@doi [\aap] {10.1051/0004-6361/201526179}, \href
  {https://ui.adsabs.harvard.edu/abs/2015A&A...579L...8X} {579, L8}

\bibitem[\protect\citeauthoryear{{Xu}, {Dufour}, {Klein}, {Melis}, {Monson},
  {Zuckerman}, {Young}  \& {Jura}}{{Xu} et~al.}{2019}]{Xu2019}
{Xu} S.,  {Dufour} P.,  {Klein} B.,  {Melis} C.,  {Monson} N.~N.,  {Zuckerman}
  B.,  {Young} E.~D.,   {Jura} M.~A.,  2019, \mn@doi [\aj]
  {10.3847/1538-3881/ab4cee}, \href
  {https://ui.adsabs.harvard.edu/abs/2019AJ....158..242X} {158, 242}

\bibitem[\protect\citeauthoryear{{Zinnecker} \& {Friedrich}}{{Zinnecker} \&
  {Friedrich}}{2001}]{Zinnecker01}
{Zinnecker} H.,  {Friedrich} S.,  2001, in {Schielicke} E.~R.,  ed.,
  Astronomische Gesellschaft Meeting Abstracts Vol. 18, Astronomische
  Gesellschaft Meeting Abstracts.

\bibitem[\protect\citeauthoryear{{Zinnecker} \& {Kitsionas}}{{Zinnecker} \&
  {Kitsionas}}{2008}]{Zinnecker08}
{Zinnecker} H.,  {Kitsionas} S.,  2008, in {Fischer} D.,  {Rasio} F.~A.,
  {Thorsett} S.~E.,   {Wolszczan} A.,  eds,  Astronomical Society of the
  Pacific Conference Series Vol. 398, Extreme Solar Systems. p.~155

\bibitem[\protect\citeauthoryear{{Zinnecker}, {Correia}, {Brandner},
  {Friedrich}  \& {McCaughrean}}{{Zinnecker} et~al.}{2006}]{Zinnecker06}
{Zinnecker} H.,  {Correia} S.,  {Brandner} W.,  {Friedrich} S.,   {McCaughrean}
  M.,  2006, in {Aime} C.,  {Vakili} F.,  eds, IAU Colloq. 200: Direct Imaging
  of Exoplanets: Science \& Techniques. pp 19--24,
  \mn@doi{10.1017/S1743921306009021}

\bibitem[\protect\citeauthoryear{{Zuckerman} \& {Becklin}}{{Zuckerman} \&
  {Becklin}}{1987}]{Zuckerman87}
{Zuckerman} B.,  {Becklin} E.~E.,  1987, \mn@doi [\apjl] {10.1086/184962},
  \href {http://adsabs.harvard.edu/abs/1987ApJ...319L..99Z} {319, L99}

\bibitem[\protect\citeauthoryear{{Zuckerman}, {Melis}, {Klein}, {Koester}  \&
  {Jura}}{{Zuckerman} et~al.}{2010}]{Zuckerman2010}
{Zuckerman} B.,  {Melis} C.,  {Klein} B.,  {Koester} D.,   {Jura} M.,  2010,
  \mn@doi [\apj] {10.1088/0004-637X/722/1/725}, \href
  {https://ui.adsabs.harvard.edu/abs/2010ApJ...722..725Z} {722, 725}

\bibitem[\protect\citeauthoryear{{Zuckerman}, {Xu}, {Klein}  \&
  {Jura}}{{Zuckerman} et~al.}{2013}]{Zuckerman13}
{Zuckerman} B.,  {Xu} S.,  {Klein} B.,   {Jura} M.,  2013, \mn@doi [ApJ]
  {10.1088/0004-637X/770/2/140}, \href
  {http://adsabs.harvard.edu/abs/2013ApJ...770..140Z} {770, 140}

\bibitem[\protect\citeauthoryear{{van Saders} \& {Gaudi}}{{van Saders} \&
  {Gaudi}}{2011}]{vanSaders11}
{van Saders} J.~L.,  {Gaudi} B.~S.,  2011, \mn@doi [ApJ]
  {10.1088/0004-637X/729/1/63}, \href
  {http://adsabs.harvard.edu/abs/2011ApJ...729...63V} {729, 63}

\bibitem[\protect\citeauthoryear{{von Hippel}}{{von
  Hippel}}{1998}]{vonHippel98}
{von Hippel} T.,  1998, \mn@doi [\aj] {10.1086/300296}, \href
  {https://ui.adsabs.harvard.edu/abs/1998AJ....115.1536V} {115, 1536}

\makeatother
\end{thebibliography}


\bsp	
\label{lastpage}
\end{document}